\documentclass[fleqn,usenatbib]{mnras}

\usepackage{graphicx}	
\usepackage{xcolor}
\usepackage{amsmath}	
\usepackage{xspace}  
\usepackage{amsmath}
\usepackage{CJKutf8}
\usepackage{multirow}  
\usepackage{tabularx}
\usepackage{makecell}
\usepackage{supertabular,booktabs}

\newcommand{\zzmgsns}{\textit{z}0MGS}
\newcommand{\zzmgs}{\zzmgsns\xspace}
\newcommand{\herschel}{\textit{Herschel}\xspace}
\newcommand{\zzmgsherschel}{\zzmgsns-\herschel}

\newcommand{\mASIAA}{Institute of Astronomy and Astrophysics, Academia Sinica, No. 1, Sec. 4, Roosevelt Road, Taipei 106216, Taiwan}

\newcommand{\mCASS}{Center for Astrophysics and Space Sciences, Department of Physics, University of California, San Diego, 9500 Gilman Drive, La Jolla, CA 92093, USA}

\newcommand{\mUCSD}{Department of Astronomy \& Astrophysics, University of California, San Diego, 9500 Gilman Drive, La Jolla, CA 92093, USA}

\newcommand{\mCFA}{Center for Astrophysics $|$ Harvard \& Smithsonian, 60 Garden Street, Cambridge, MA 02138, USA}

\newcommand{\mUMD}{Department of Astronomy, University of Maryland, 4296 Stadium Drive, College Park, MD 20742, USA}

\newcommand{\mOSU}{Department of Astronomy, The Ohio State University, 4055 McPherson Laboratory, 140 West 18th Avenue, Columbus, OH 43210, USA}

\newcommand{\mCCAPP}{Center for Cosmology and Astroparticle Physics, 191 West Woodruff Avenue, Columbus, OH 43210, USA}

\newcommand{\mUGENT}{Sterrenkundig Observatorium, Universiteit Gent, Krijgslaan 281 S9, B-9000 Gent, Belgium}

\newcommand{\mOxford}{Sub-department of Astrophysics, Department of Physics, University of Oxford, Keble Road, Oxford OX1 3RH, UK}

\newcommand{\mOsaka}{Theoretical Astrophysics, Department of Earth and Space Science, Osaka University, 1-1 Machikaneyama, Toyonaka, Osaka 560-0043, Japan}

\ifdefined\equationautorefname
    \renewcommand{\equationautorefname}{Eq.}
\fi
\ifdefined\sectionautorefname
    \renewcommand{\sectionautorefname}{Section}
\fi
\ifdefined\subsectionautorefname
    \renewcommand{\subsectionautorefname}{Section}
\fi
\ifdefined\subsubsectionautorefname
    \renewcommand{\subsubsectionautorefname}{Section}
\fi

\newcommand{\SigmasfrUnit}{\ensuremath{{\rm M_\odot\,yr}^{-1}\,{\rm kpc}^{-2}}\xspace}

\newcommand{\SEDUnit}{{\rm MJy\,sr\textsuperscript{-1}}\xspace}
\newcommand{\acoUnit}{\ensuremath{{\rm M_\odot\,pc^{-2}\,(K\,km\,s^{-1})^{-1}}}\xspace}

\newcommand{\metal}{12+\logt({\rm O/H})\xspace}
\newcommand{\CO}{{\rm CO}\xspace}

\newcommand{\HI}{\textsc{H\,i}\xspace}
\newcommand{\HII}{\textsc{H\,ii}\xspace}

\newcommand{\Sigmad}{\ensuremath{\Sigma_\mathrm{dust}}\xspace}
\newcommand{\Sigmastar}{\ensuremath{\Sigma_\star}\xspace}
\newcommand{\Sigmasfr}{\ensuremath{\Sigma_\mathrm{SFR}}\xspace}
\newcommand{\Sigmagas}{\ensuremath{\Sigma_\mathrm{gas}}\xspace}

\newcommand{\Sigmaatom}{\ensuremath{\Sigma_\mathrm{atom}}\xspace}
\newcommand{\Sigmamol}{\ensuremath{\Sigma_\mathrm{mol}}\xspace}

\newcommand{\aco}{\ensuremath{\alpha_\CO}\xspace}

\newcommand{\Ubar}{\ensuremath{\overline{U}}\xspace}
\newcommand{\Umin}{\ensuremath{U_\mathrm{min}}\xspace}
\newcommand{\qpah}{\ensuremath{q_\mathrm{PAH}}\xspace}

\newcommand{\shortminus}{\scalebox{0.75}[1.0]{\ensuremath{-}}}
\newcommand{\coone}{\textup{CO}\,\ensuremath{(1{\shortminus}0)}\xspace}
\newcommand{\cotwo}{\textup{CO}\,\ensuremath{(2{\shortminus}1)}\xspace}

\renewcommand{\metal}{\ensuremath{12+\log({\rm O/H})}\xspace}

\newcommand{\arcdeg}{\ensuremath{^{\prime}}}
\newcommand{\nodata}{\ensuremath{~\cdots~}}

\usepackage{threeparttable}

\usepackage{newtxtext,newtxmath}

\usepackage[T1]{fontenc}

\DeclareRobustCommand{\VAN}[3]{#2}
\let\VANthebibliography\thebibliography
\def\thebibliography{\DeclareRobustCommand{\VAN}[3]{##3}\VANthebibliography}

\usepackage{graphicx}	
\usepackage{amsmath}	

\defcitealias{Chiang24_alphaCO_main}{C24}
\defcitealias{Hirashita23b_aCO}{H23}

\title[$\alpha_{\rm CO}$ and grain size distribution]{CO-to-H$_2$ conversion factor and grain size distribution through the analysis of $\alpha_\mathrm{CO}$--$q_\mathrm{PAH}$ relation}

\author[I-Da Chiang et al.]{
I-Da Chiang \begin{CJK*}{UTF8}{bkai}(江宜達)\end{CJK*},$^{1}$\thanks{E-mail: idchiang@asiaa.sinica.edu.tw}  
Hiroyuki Hirashita,$^{1,2}$  
J\'er\'emy Chastenet,$^{3}$  
Karin M. Sandstrom,$^{4}$  
Eric W. Koch,$^{5}$  
\newauthor
Adam K. Leroy,$^{6,7}$  
Yu-Hsuan Teng,$^{8,9}$  
and
Thomas G. Williams$^{10}$  
\\
$^{1}$\mASIAA\\
$^{2}$\mOsaka\\
$^{3}$\mUGENT\\
$^{4}$\mUCSD\\
$^{5}$\mCFA\\
$^{6}$\mOSU\\
$^{7}$\mCCAPP\\
$^{8}$\mCASS\\
$^{9}$\mUMD\\
$^{10}$\mOxford
}

\date{Accepted XXX. Received YYY; in original form ZZZ}

\pubyear{2024}

\begin{document}
\label{firstpage}
\pagerange{\pageref{firstpage}--\pageref{lastpage}}
\maketitle

\begin{abstract}
The CO-to-H$_2$ conversion factor ($\alpha_\mathrm{CO}$) is expected to vary with dust abundance and grain size distribution through the efficiency of shielding gas from CO-dissociation radiation.
We present a comprehensive analysis of $\alpha_\mathrm{CO}$ and grain size distribution for nearby galaxies, using the PAH fraction ($q_\mathrm{PAH}$) as an observable proxy of grain size distribution.
We adopt the resolved observations at 2~kpc resolution in 42 nearby galaxies, where $\alpha_\mathrm{CO}$ is derived from measured metallicity and surface densities of dust and \textsc{H~i} assuming a fixed dust-to-metals ratio. We use an analytical model for the evolution of H$_2$ and CO, in which the evolution of grain size distribution is controlled by the dense gas fraction ($\eta$).
We find that the observed level of $q_\mathrm{PAH}$ is consistent with the diffuse-gas-dominated model ($\eta=0.2$) where dust shattering is more efficient. Meanwhile, the slight decreasing trend of observed $q_\mathrm{PAH}$ with metallicity is more consistent with high-$\eta$ predictions, likely due to the more efficient loss of PAHs by coagulation.
We discuss how grain size distribution (indicated by $q_\mathrm{PAH}$) and metallicity impact $\alpha_\mathrm{CO}$; we however did not obtain conclusive evidence that the grain size distribution affects $\alpha_\mathrm{CO}$. Observations and model predictions show similar anti-correlation between $\alpha_\mathrm{CO}$ and 12+log(O/H).
Meanwhile, there is a considerable difference in how resolved $\alpha_\mathrm{CO}$ behaves with $q_\mathrm{PAH}$. The observed $\alpha_\mathrm{CO}$ has a positive correlation with $q_\mathrm{PAH}$, while the model-predicted $\alpha_\mathrm{CO}$ does not have a definite correlation with $q_\mathrm{PAH}$. This difference is likely due to the limitation of one-zone treatment in the model.
\end{abstract}




\section{Introduction}

Stars form in cold, dense molecular clouds in the interstellar medium (ISM). Molecular gas fuels star formation, and is a key factor diagnosing star-forming conditions. Therefore, observing molecular gas is essential for understanding star formation and galaxy evolution. However, the major component of molecular gas, H$_2$, does not possess a permanent dipole moment and thus does not emit efficiently in cold molecular clouds.
As a result, observers often use the low-$J$ ($J$ is the rotational energy level) emission lines of the second most abundant molecule, CO, to trace the molecular gas.

The standard practice for astronomers is to convert the observed \coone integrated intensity at wavelength $\lambda =1.3$~mm ($W_\mathrm{CO}$~[K~km~s$^{-1}$]) to molecular gas surface density (\Sigmamol~[M$_{\sun}$~pc$^{-2}$]) or H$_2$ column density ($N\mathrm{(H_2)}$~[cm$^{-2}$]) as follows:
\begin{equation}\label{eq:aco}
    \Sigmamol = \aco W_\mathrm{CO},~\mathrm{or}~N\mathrm{(H_2)}=X_\mathrm{CO}W_\mathrm{CO},
\end{equation}
where \aco and $X_\mathrm{CO}$ are the ``CO-to-H$_2$ conversion factor'' under different conventions. The conventional CO-to-H$_2$ conversion factor applicable to the Milky Way (MW) environment is $\aco=4.35~\acoUnit$ \citep[see][]{BOLATTO13}\footnote{Although \aco varies within molecular clouds, it provides a reasonable estimate for the integrated intensity of molecular clouds in the MW \citep{Sofue20}.}, which corresponds to $X_\mathrm{CO}=2\times 10^{20}~\mathrm{cm^{-2}~(K~km~s^{-1})^{-1}}$.\footnote{The \aco (\Sigmamol) convention includes a 1.36 factor accounting for helium mass while the $X_\mathrm{CO}$ ($N\mathrm{(H_2)}$) convention does not.}
In this paper, we follow the \aco (or \Sigmamol) convention.

The CO-to-H$_2$ conversion factor is environment-dependent.
Obtaining appropriate values of \aco for various environments is crucial to accurately constrain the initial conditions for star formation. Astronomers have found two major trends that set the value of \aco.
First, \aco tends to drop in galaxy centres, star-forming regions and (ultra-)luminous infrared (IR) galaxies \citep{DownesSolomonRadford93,DownesSolomon98,Israel09a,Israel09b,Israel20,Weiss01,Papadopoulos12,SANDSTROM13,Herrero-Illana19,Jiao21,Teng22,Teng23,DenBrok23,Chiang24_alphaCO_main}. This is interpreted as \aco decreasing with the rise of the CO emissivity, which increases with gas density, temperature and optical depth.
Secondly, \aco rises at low metallicity and low dust-to-gas ratio (D/G) environments. This is due to decreased shielding of CO-dissociating radiation and thus decreased CO-emitting area in molecular clouds, which is often phrased as `CO-dark' gas \citep{Arimoto96,ISRAEL97,Papadopoulos02_DarkGas,Grenier05_DarkGas,WOLFIRE10,LEROY11,PlanckCollaborationXIX2011,HUNT15,ACCURSO17,Madden20_DarkGas}.
Some of the latest observation-based formulae tracing the above mechanisms are summarized in the \citet{SchinnererLeroy24_SFReview} review \citep[see also][]{BOLATTO13}.

It is expected that the second effect---increased \aco at low metallicity---has a direct link to galaxy evolution through metal enrichment.
The CO-dark gas is tightly linked to shielding of CO-dissociating radiation by gas and dust \citep[e.g.][]{Lee96_CO_Dissociation,GloverMacLow11}. Given the local physical conditions, \aco can in principle be derived theoretically by calculating the formation and destruction of H$_2$ and CO \citep[e.g.][]{NARAYANAN11,GloverMacLow11,FELDMANN12}. 
Most existing formulae use a single parameter, i.e.\ metallicity or D/G, to trace the dust shielding. Recent modelling further investigated the effect of dust properties, especially focusing on the grain size distribution\footnote{By ``grain size distribution'', we refer to the size distribution of all types of dust grains, including PAHs.} \citep[][hereafter H23]{HIRASHITA17_HH17_XCO,Hirashita23b_aCO}. These models show that even if the dust abundance is the same, different grain size distributions predict different shielding efficiencies of ultraviolet (UV) dissociating radiation as it changes the crosssections.

Based on the analytic approach of representing the grain size distribution at two grain radii by \citet{HIRASHITA17_HH17_XCO}, \citet{Chen18_h2_co_simulation} post-processed a disc-galaxy simulation, and found that the difference in grain size distribution causes an appreciable change of the CO abundance. A grain size distribution with a higher portion of small grains tends to result in a higher CO abundance since smaller grains absorb UV radiation more efficiently at fixed dust abundance. This conclusion was confirmed with a full treatment of grain size distribution (without approximating it by two sizes) by \citetalias{Hirashita23b_aCO}, who showed that the evolution of grain size distribution regulated by the dense-gas fraction causes nearly an order-of-magnitude difference in \aco at sub-solar to solar metallicities.
Therefore, the effect of grain size distribution should be quantitatively checked along with the interpretation of the metallicity dependence of \aco.

In this paper, we examine the effect of grain size distribution on \aco using observations. Given that \aco is affected by the local physical conditions, we adopt the latest spatially resolved measurements of nearby galaxies by \citet[hereafter C24]{Chiang24_alphaCO_main}. \citetalias{Chiang24_alphaCO_main} compiled and analyzed multi-wavelength data for dust properties, neutral gas emission lines, and auxiliary information from $\sim 40$ nearby galaxies. They derived \aco from the surface densities of dust and \HI and metallicity, assuming a constant dust-to-metals ratio (D/M). This results in $\sim 1400$ independent measurements of \aco across various environments in nearby star-forming galaxies at a uniform physical resolution of 2\,kpc. This dataset provides a testing ground for understanding how \aco evolves with local physical conditions, especially dust properties.

An observational proxy for the grain size distribution is necessary for the above goal. Specifically, it is possible to extract information on the grain size distribution from the widely available IR data, e.g.\ from \textit{Herschel Space Observatory} \citep[\textit{Herschel},][]{PILBRATT10} far-IR and Wide-field Infrared Survey Explorer \citep[WISE,][]{WRIGHT10} mid-IR data.
Since the \citetalias{Chiang24_alphaCO_main} sample has abundant IR data, it also represents an ideal sample to extract information on the grain size distribution from the dust emission spectral energy distributions (SEDs).
Polycyclic aromatic hydrocarbons (PAHs)\footnote{We treat PAHs as one of the dust species. Thus, the calculated grain size distributions in this paper include PAHs at small grain radii.} are particularly useful as an observational tracer of small grains, since they have prominent features at mid-IR wavelengths. Meanwhile, the total dust abundance can be evaluated by analyzing the far-IR SED, allowing the fraction of PAH to total dust mass (\qpah) to be measured and then used as an observational proxy for the fraction of small grains, which in turn has a direct link to the grain size distribution. Indeed, \citet{Matsumoto24} used \qpah as an indicator of grain size distribution in their simulations and showed that the model-predicted \qpah has a direct correspondence with the observed mid-IR PAH emission luminosity divided by the far-IR thermal emission. Therefore, \qpah can potentially bridge the models and observations by acting as an indicator of grain size distribution.

In \citetalias{Chiang24_alphaCO_main}, \qpah is one of the quantities obtained by fitting the \citet{DRAINE07} physical dust model to the WISE mid-IR and \textit{Herschel} far-IR photometry data. The fitting is performed in \citet{Chastenet24_Herschel}, following the same methodology reported in \citet{Chastenet21_M101}, and this is the same IR SED fitting used to estimate the dust surface density and so \aco. This means that \qpah is immediately available as an indicator of grain size distribution for all \citetalias{Chiang24_alphaCO_main} targets and we use these \qpah estimates as the default tracer for grain size distribution in the analysis.
We also examine the \qpah predicted by \citet{Hirashita23a_SizeDistr}, based on a model designed to reproduce the MW \qpah and grain size distribution at solar metallicity. 

This paper is organized as follows. In \autoref{sec:data}, we review the \citetalias{Chiang24_alphaCO_main} observations, including the explanations of how \aco and dust properties are measured and derived. In \autoref{sec:model}, we give an overview of the \citetalias{Hirashita23b_aCO} model.
In \autoref{sec:results}, we present the observed relations among \qpah, \aco and metallicity, and how these trends compare to the model prediction. 
In \autoref{sec:discussion}, we discuss the interpretation of our results and other possible tracers for grain size distribution.
Finally, we summarize our findings in \autoref{sec:summary}.

\section{Data}\label{sec:data}
In this work, we utilize the spatially resolved measurements of \aco and other relevant physical quantities presented in \citetalias{Chiang24_alphaCO_main}. Here, we briefly describe the dataset and refer the reader to \citetalias{Chiang24_alphaCO_main} for more details. 

\citetalias{Chiang24_alphaCO_main} assembled multi-wavelength data relevant for estimating the surface densities of various components, i.e.\ dust, neutral gas, stellar mass, SFR, and metallicity, in nearby galaxies. We compile data from 42 galaxies in our analysis, including the 37 presented in \citetalias{Chiang24_alphaCO_main} and 5 additional ones. All the data are convolved to 2\,kpc physical resolution ($\geq 21\arcsec$ angular resolution, depending on the distance) with pixel sizes of 2/3\,kpc. We list the properties of all galaxies and references of adopted data in Appendix \ref{app:samples}. All galaxies have distances within $20\,\mathrm{Mpc}$ such that the lowest resolution data (dust or \HI with $\sim 21\arcsec$ resolution) can be analyzed at a uniform 2\,kpc resolution. 
The target galaxies are selected to have spatially resolved observation of neutral gas emission lines (\HI 21\,cm and either \coone at 115~GHz or \cotwo at 230~GHz) and maps from the $z$0MGS-\textit{Herschel}/Dust catalogue \citep{Chastenet24_Herschel}. The $z$0MGS catalogue also includes auxiliary data in the mid-IR and UV, which trace surface densities of star formation rate (SFR) and stellar mass (denoted as \Sigmasfr and \Sigmastar, respectively; \citealt{LEROY19}). 
To the best of our knowledge, this is the largest spatially resolved dataset with \aco, \qpah, and metallicity.
Below we briefly describe the quantities used in this paper and clarify some differences from \citetalias{Chiang24_alphaCO_main}.

\noindent\textbf{CO-to-H$_2$ conversion factor.} \citetalias{Chiang24_alphaCO_main} measured \aco using a dust-based strategy. The key assumption in their method is a fixed value for the fraction of metals locked in the solid phase, i.e.\ D/M. They calculated the molecular gas surface density (\Sigmamol) from the measured values of dust surface density (\Sigmad), atomic gas surface density (from \HI), and metallicity under the assumed D/M. Then they divided the calculated \Sigmamol by integrated CO intensity ($I_\mathrm{CO}$) to evaluate \aco. The value of D/M adopted by \citetalias{Chiang24_alphaCO_main} is 0.55; however, we adopt the value of 0.48 from \citetalias{Hirashita23b_aCO} as the fiducial case in this work for consistency with the model (more details in \autoref{sec:model}). This change in D/M causes \aco to increase about 0.07 to 0.10\,dex from the \citetalias{Chiang24_alphaCO_main} values.\footnote{According to \citetalias{Chiang24_alphaCO_main}, the reasonable range of D/M of our sample is roughly 0.4--0.7, which corresponds to $\sim$0.1--0.2~dex systematical shift in \aco from their values in this work.} We note that there are other strategies for converting dust observations to hydrogen mass, e.g.\ using a constant D/G \citep[e.g.][]{Boulanger96}, minimizing the scatter in D/G \citep[e.g.][]{SANDSTROM13}, or nonlinear dust optical depth to \Sigmagas conversion \citep[e.g.][]{Okamoto17,Hayashi19}. We refer the readers to \citet{BOLATTO13} and \citetalias{Chiang24_alphaCO_main} for the discussion of these methodologies.

\noindent\textbf{Dust properties.} The dust properties from the $z$0MGS-\textit{Herschel}/Dust catalogue are derived by fitting the observed dust emission SED with the \citet{DRAINE07} dust model. The details of the IR data processing and dust SED fitting are reported in \citet{Chastenet24_Herschel}. The IR SED used in the fitting includes the WISE 3.4, 4.6, 12, and 22\,\micron\ bands, and the \textit{Hershel} 70, 100, 160, and 250\,\micron\ bands. These IR maps are first convolved to circular Gaussian point spread functions at the final physical resolution (2\,kpc) before the fitting.
The dust SEDs are fitted using the \citet{DRAINE07} physical dust model with dust opacity correction factor derived in \citet{Chastenet21_M101}, applied through the \texttt{DustBFF} tool, which considers the full covariance matrix between photometry bands \citep{GORDON14}.
The product of fitting includes maps of the dust mass surface density (\Sigmad), the interstellar radiation field (ISRF) properties (\Umin, $\gamma$ and \Ubar), and the fractional mass of PAHs to total dust (PAH fraction, \qpah). We define the PAHs as small aromatic grains with radii $a<13$\,\AA\ \citep[corresponding to $N_C \leq 10^3$ C atoms. see Equation\,3 and Section\,10.3 of][]{DRAINE07} in both the SED fitting and the model calculations.

A specific caveat of using the \citet{Chastenet24_Herschel} dust properties in this study is that the grain size distribution is not explicitly fitted to the observations. Instead, they fit \qpah and the grain size distribution almost solely varies with the inferred PAH abundance. This is different from the \citetalias{Hirashita23b_aCO} model, where \qpah is not the only factor that determines the functional form of grain size distribution.
Practically, we expect that \qpah and other dust properties derived from the SED fitting are not sensitive to the detailed assumptions regarding the grain size distribution. The evaluation of the total dust mass is dominated by the far-IR part of the SED, which is robust against the change of the grain size distribution, while the evaluation of the PAH mass is based on the level of the prominent PAH features. Therefore, \qpah is practically determined by the PAH emission strength relative to the FIR luminosity, which is not affected by the detailed functional shape of the grain size distribution.  

\noindent\textbf{Metallicity.} \citetalias{Chiang24_alphaCO_main} used oxygen abundance, \metal, to trace the metallicity ($Z$). They assumed a fixed oxygen-to-total-metal mass ratio and
converted \metal to $Z$ with $Z = 0.0134 \times 10^{\metal - 8.69}$, where 8.69 is the adopted solar oxygen abundance \citep[$\metal_\odot$,][]{ASPLUND09}. To calculate \metal for each pixel, \citetalias{Chiang24_alphaCO_main} adopted the radial gradient of \metal with the PG16S calibration \citep{PilyuginGrebel16} from the PHANGS-MUSE survey \citep{Emsellem22_PHANGS-MUSE,Groves23} and the \citet{Zurita21} compilation. For galaxies without measurement from either dataset, \citetalias{Chiang24_alphaCO_main} adopted an empirical formula: $\metal=8.56 + 0.01xe^{-x}-0.1r_\mathrm{g}/R_\mathrm{e}$, where $x=\log(M_\star/\mathrm{M}_\odot) - 11.5$, $M_\star$ is the total stellar mass of the galaxy, $r_\mathrm{g}$ is the galactocentric radius and $R_\mathrm{e}$ is the effective radius. This is based on the two-step strategy developed in \citet{Sun20}, where we estimate the metallicity at 1 $R_\mathrm{e}$ from $M_\star$ \citep[see][]{Sanchez19} and apply a universal gradient \citep[see][]{Sanchez14}. We refer the readers to \citetalias{Chiang24_alphaCO_main} for the derivation and possible caveats of the methodology.

\noindent\textbf{SFR.} \citetalias{Chiang24_alphaCO_main} traced the SFR surface density (\Sigmasfr) using the \zzmgs data \citep{LEROY19} and conversion formula presented in \citet{Belfiore23}. They utilized the \zzmgs compilation of the background-subtracted intensities of the WISE 
$\lambda\sim 22$\,\micron\ (hereafter WISE4) data and the Galaxy Evolution Explorer \citep[GALEX,][]{MARTIN05} $\lambda\sim154\,\rm nm$ (hereafter FUV) data, denoted as $I_\mathrm{WISE4}$ and $I_\mathrm{FUV}$, respectively. The conversion is: $\frac{\Sigmasfr}{1\,\SigmasfrUnit} = 
8.85 \times 10^{-2} \frac{I_{\rm FUV}}{1\,\SEDUnit} + 3.02 \times 10^{-3} \frac{I_{\rm WISE4}}{1\,\SEDUnit}$.

\noindent\textbf{Signal-to-noise ratio (S/N) cut.} Following \citetalias{Chiang24_alphaCO_main}, we constrain our fiducial sample used for pixel-by-pixel analysis to pixels with $\textrm{S/N}\geq 1$ for both derived \Sigmamol and observed integrated CO intensity ($W_\mathrm{CO}$). Note that in \citetalias{Chiang24_alphaCO_main} (see their Eq. 4), \Sigmamol is derived from \HI, metallicity, and IR photometry data for deriving \Sigmad. Consequently, the uncertainty of \Sigmamol is propagated from \HI, metallicity, and IR data. We additionally impose $\textrm{S/N}\geq 3$ for IR data as recommended in \citet{Chastenet21_M101} since we use dust data alone in a few analyses.

\noindent\textbf{Completeness.} We calculate several statistical relations (e.g.\ correlations and linear regression) between \aco and relevant quantities. To avoid selection bias, we only use a subsample that is \textit{complete} for the target quantity to examine the statistical properties. Following \citetalias{Chiang24_alphaCO_main}, a \textit{complete} sample is defined as satisfying \textit{completeness}\,$\geq$\,50\%, and the \textit{completeness} is the ratio of pixels that satisfy the S/N cuts mentioned above to the total number of pixels in a bin of the target quantity. For example, \metal is complete in the range of $\sim$8.4 to $\sim$8.7.

\noindent\textbf{Differences from \citetalias{Chiang24_alphaCO_main}.} Here we summarize the differences between the measurements in this work and \citetalias{Chiang24_alphaCO_main}. 
First, we use $\mathrm{D/M}=0.48$ as the fiducial case in this work instead of 0.55 in \citetalias{Chiang24_alphaCO_main}. This change was made for consistency with the calculations in the adopted model (\autoref{sec:model}). 
Second, we analyze 5 additional galaxies, which now meet our S/N cut thanks to the adoption of the new fiducial D/M (previously they missed the S/N cut in \Sigmamol). We list the properties of these galaxies in \autoref{app:samples}.
Third, for metallicity derived from the \citetalias{Chiang24_alphaCO_main} empirical formula, we constrain the radii to $0.3 \leq r_\mathrm{g}/R_\mathrm{e} \leq 2.0$ as recommended by \citet{Sanchez14}.
Fourth, we use \coone data whenever possible instead of presenting \coone and \cotwo in parallel as \citetalias{Hirashita23b_aCO} focused on \coone. In galaxies where \coone is unavailable, we use \cotwo data with the \Sigmasfr-dependent line ratio suggested by \citet{SchinnererLeroy24_SFReview}:
\begin{equation}
    R_{21} = 0.65\Big(\frac{\Sigmasfr}{1.8\times 10^{-2}\,\SigmasfrUnit}\Big)\,\mathrm{with~min~0.35,~max~1.0.}
\end{equation}

\section{Model}\label{sec:model}
We utilize theoretical models that describe \aco in a manner consistent with the dust life cycle. In particular, we focus on the evolution of grain size distribution as done by \citetalias{Hirashita23b_aCO}. Since the PAH fraction (\qpah) is used as an observational proxy of grain size distribution, it is also important to adopt models that are capable of predicting \qpah. Below we briefly review the model used in this paper and refer the interested reader to \citetalias{Hirashita23b_aCO} for details. We also apply some modifications to the evolution of grain size distribution based on \citet{Hirashita23a_SizeDistr}.

The evolution model of grain size distribution is taken from \citet{Hirashita_Murga20}, which is developed based on \citet{Asano13_GrainSize} and \citet{HIRASHITA+AOYAMA19}.
We calculate the dust enrichment by stars (SNe and AGB stars) in a manner consistent with the metal enrichment and assume a log-normal grain size distribution (with a typical grain radius of 0.1\,$\micron$) for the stellar dust sources. We adopt a star formation time-scale of 5\,Gyr, which regulates the metal enrichment.
The star formation time scale has only a minor influence on the results as long as we use the metallicity (not the age) to measure the evolutionary stage (as done in \autoref{sec:results}).
Thus, we simply fix the star formation time-scale to 5\,Gyr, which is broadly appropriate for normally star-forming galaxies (like the MW and typical spiral galaxies) similar to our observational sample.

We consider interstellar processing of dust: dust destruction by SN shocks, accretion of gas-phase metals in the dense ISM, coagulation in the dense ISM, and shattering in the diffuse ISM (the basic equations for these processes are summarized by \citealt{HIRASHITA+AOYAMA19}). Given that our model does not account for the hydrodynamical evolution of the ISM, we treat the mass fraction of the dense ISM, $\eta$, as a fixed parameter.
In the metallicity range of interest in this work, the role of $\eta$ is mainly to regulate the balance between shattering and coagulation. A larger value of $\eta$ tends to produce fewer small grains because coagulation (shattering) becomes stronger (weaker).

The grain species are treated based on \citet{Hirashita_Murga20}.
The calculated grain size distribution is decomposed into silicate and carbonaceous dust based on the abundance ratio between Si and C in the ISM. The carbonaceous species is further divided into aromatic and nonaromatic grains considering aromatization and aliphatization. Since aromatization predominantly occurs in the diffuse ISM, the resulting aromatic fraction is approximately $1-\eta$. We also investigate a case where the aromatic fraction is unity in order to present a maximally allowed \qpah. This maximum value is realized in the case where aliphatization is neglected.

We also investigate a model in which the PAH abundance is enhanced following the prescription in \citet{Hirashita23a_SizeDistr}. This is motivated by the underpredictions of PAH emission in the above-mentioned models. This is referred to as the \textit{enPAH} model (meaning the model with PAH enhancement), while the model without this prescription is referred to as the \textit{standard} model. In the enPAH model, we hypothesize that
small carbonaceous grains, once they are formed mainly by shattering, remain unprocessed by coagulation, accretion, and shattering. In particular, decoupling from the coagulation process is the most essential point, since it avoids small carbonaceous grains (including PAHs) being attached onto larger grains.

For both (standard and enPAH) models, we set the maximum dust-to-metal ratio (D/M); that is, the increase of D/M by accretion is suppressed if it would exceed 0.48 \citepalias{Hirashita23b_aCO}. This is based on the consideration that not all metals are available for dust, and is widely consistent with the dust-to-metal ratio in the MW and nearby galaxy discs \citep{ISSA90,LEROY11,DRAINE14,Vilchez19,Chiang21}. In the metallicity range of interest in this paper ($\sim 0.5$--$1.0\,Z_\odot$), D/M is broadly saturated to 0.48 and is consistent with the assumption of constant D/M adopted by \citetalias{Chiang24_alphaCO_main}.

The theoretical calculation of \aco is based on \citetalias{Hirashita23b_aCO}, whose model is extended from \citet{HIRASHITA17_HH17_XCO} to include the grain size distribution. They considered a typical molecular cloud, in which the shielding of ISRF and the formation of H$_2$ on the grain surfaces are treated in a manner consistent with the calculated grain size distribution. To quantify the effect of grain size distribution, we define the ratio between the total grain surface area and the total dust mass and refer to it as the SD ratio (SDR). A larger SDR means that the grain size distribution is biased to small sizes.

The physical environments considered in the model do not perfectly align with those sampled by the observations. To address this, we constrain the environments where we compare the model predictions to the observations. We set the constraint according to the D/G-to-metallicity relation, which is a key parameter for both the observed and model-predicted data set. Specifically, we use $\text{D/M}=0.48$ for deriving \aco in the observations and only compare them to model predictions that yield D/M between 0.48 and 0.43 (90\% of 0.48).

\section{Results}\label{sec:results}

We first present the general behaviour of \qpah, which is used as an indicator of grain size distribution in this paper. Then, we investigate the relation between \qpah and \aco to understand how the evolution of grain size distribution affects the CO-to-H$_2$ conversion factor. We use the metallicity (indicated by \metal) as an indicator of the evolutionary stage for both models and observations.

\subsection{Properties of \texorpdfstring{\qpah}{qPAH}}\label{sec:results:qpah}

\begin{figure}
    \centering
	\includegraphics[width=0.99\columnwidth]{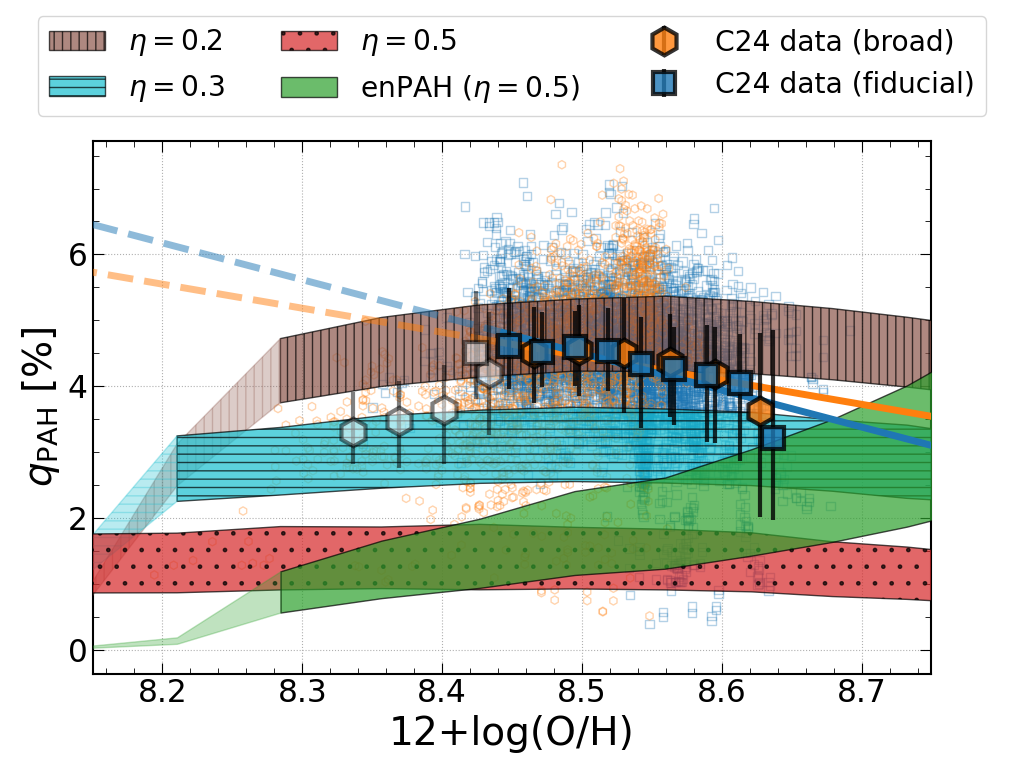}
    \caption{
    \qpah as a function of metallicity.
    The filled blue squares (orange hexagons) show the binned observed data in the fiducial (broad) sample and the lines in the same colour show the linear regression of the corresponding sample.
    The linear regression only utilizes measurements in the metallicity range where data are complete. The unfilled big hexagons/squares and the dashed lines indicate the region where observed data is not complete. 
    The unfilled small blue squares (orange hexagons) show the pixel-by-pixel observed data in the fiducial (broad) sample.
    The shaded regions show the model-predicted \qpah, where the lower limit comes from the standard aromatic fraction in \citetalias{Hirashita23b_aCO} and the upper limit is set by an aromatic fraction of 1.0 (corresponding to the maximum PAH prescription). The transparent shaded regions indicate the environment where D/M is less than 90 per cent of the fiducial value (0.48, see \autoref{sec:model}).
    }
    \label{fig:qpah-metal}
\end{figure}

We first show how \qpah evolves with metallicity in observations. The median of the measured \qpah is 4.49 per cent, whereas the values at 16th--84th percentiles are 3.49--5.36 per cent.
As shown in \autoref{fig:qpah-metal}, we observe a weak anti-correlation between \qpah and metallicity (the blue data points). The negative \qpah--metallicity slope is steeper toward higher metallicity.
Meanwhile, since \aco is not involved in this analysis, we also measure the \qpah--metallicity relation with a sample with less strict S/N cut; that is, we only require $\textrm{S/N}\geq 3$ for the IR data used for dust SED fitting and neglect the S/N for $W_\mathrm{CO}$, \HI, and metallicity. 
This sample is less biased towards molecular-gas-rich (CO-bright) regions because of these relaxed S/N constraints.
We label this broadened sample as \textit{broad} in \autoref{fig:qpah-metal}. The \textit{broad} sample shows a \qpah--metallicity trend that is similar to the fiducial sample above the completeness threshold. Below the completeness cut ($\metal \lesssim 8.45$), we observe \qpah decreases toward lower metallicity. However, this trend is dominated by a few galaxies and is not taken into account in the overall statistics.

Compared to previous observations, we reproduced the trend that \qpah decreases towards lower metallicity in NGC 5457 \citep{Chastenet21_M101}, which is one of the galaxies that dominates the low-metallicity trend in \autoref{fig:qpah-metal}.
Previous galaxy-integrated observations that include low-metallicity galaxies show that \qpah increases with metallicity \citep[e.g.,][]{Draine07_SINGS,Smith07,Khramtsova14_PAH,Chastenet19_MCs,ANIANO20,Li20_PAH} when considering metallicity down to $\sim 0.1\,Z_\odot$. Since we only have complete metallicity coverage down to $\sim 0.5\,Z_\odot$, our results are not able to be thoroughly compared to their whole range measurements. In the $\sim0.5$--$1.0\,Z_\odot$ subsample, our data is consistent with the KINGFISH measurements \citep{ANIANO20}. Recent JWST observations \citep{Shivaei24_PAH} reported a roughly constant \qpah at metallicity above $0.5\,Z_\odot$ at redshift 0.7--2, which is similar to what we observed with the \textit{broad} sample. \citet{Whitcomb24PAH} recently reported that PAH relative to total IR emission is roughly constant near $Z\sim 0.6\,Z_\odot$ and drops toward both higher and lower $Z$, which is broadly consistent with our observations. 

The observed \qpah\ vs. metallicity trends are compared with the theoretical predictions in \autoref{fig:qpah-metal}. We first discuss the standard models.
At low metallicity ($\metal \lesssim 8.3$, depending on $\eta$), the standard models predict low \qpah. This is due to the lack of efficient small grain production by accretion and shattering; however, our observational data is incomplete at such low metallicity ($\metal \lesssim 8.4$) and does not provide a firm comparison.
At moderate to high metallicity ($\metal\gtrsim 8.3$, depending on $\eta$), \qpah is nearly constant in the standard models. This is because the equilibrium between shattering and coagulation for PAHs has been achieved. This is mostly the metallicity regime where robust data are available. The level of equilibrium \qpah strongly depends on $\eta$ and is higher for smaller $\eta$. Among the standard models, the one with $\eta=0.2$ has its equilibrium \qpah most consistent with the observation. This is in agreement with our previous results \citep{Hirashita_Deng_Murga20}: Lower $\eta$, which enhances the abundance of small grains (including PAHs), is favoured to reproduce the level of PAH emission \citep{Hirashita_Deng_Murga20}. This is likely consistent with the observations by \citet{Sutter24_PAH}, who found that the strength of PAH emission, relative to other small grain emission, decreases as gas density increases. They raised coagulation as a possible explanation for this decrease.
However, the standard model with low $\eta$ does not predict the observed \qpah--metallicity anti-correlation in the fiducial sample, especially at high metallicity.
This decreasing trend of \qpah with metallicity is more consistent with high-$\eta$ models where the enhanced effect of coagulation at high metallicity is more obvious. 
Due to stronger coagulation in more dust-rich (or metal-rich) regions, one would expect that \qpah decreases with metallicity (recall that PAHs are depleted by coagulation since they are attached onto larger grains).
This decline is rather consistent with high-$\eta$ predictions because of more significant coagulation towards higher metallicity.
Thus, the \qpah--metallicity trend may be in favour of high $\eta$ while the absolute level of \qpah is rather in agreement with low $\eta$. We will explore other explanations for high-metallicity \qpah decline in \autoref{sec:discussion:qpah_decline}.

As mentioned in \autoref{sec:model}, we also examine the enPAH model, which produces a larger \qpah with a larger $\eta$ $(\simeq 0.5)$ because no PAHs are removed by coagulation \citep{Hirashita23a_SizeDistr}.
However, the enPAH model produces a clearly increasing trend of \qpah as a function of metallicity, which is not supported by our spatially resolved observational data. Additionally, the absolute \qpah level is only closer to observations at high metallicity.
Previous galaxy-integrated observations do show an increasing trend of \qpah with metallicity \citep[e.g.][]{Draine07_SINGS,RemyRuyer15_PAH,GALLIANO18,ANIANO20}. However, that trend is usually more significant when comparing $Z<0.5 Z_\odot$ galaxies to $Z\sim Z_\odot$ ones instead of a smooth increment with metallicity; thus the enPAH model does not necessarily explain their findings, either. We will omit the enPAH model in the following analysis.
The unsuccessful results of the enPAH model, which decouple PAHs from interstellar processing (especially coagulation), suggest the balance between coagulation and shattering plays an important role in the flat (or decreasing) trend of \qpah with metallicity.

\subsection{\texorpdfstring{\aco}{alphaCO} and grain size distribution}\label{sec:results:alphaCO}

\begin{figure*}
    \centering
	\includegraphics[width=0.99\textwidth]{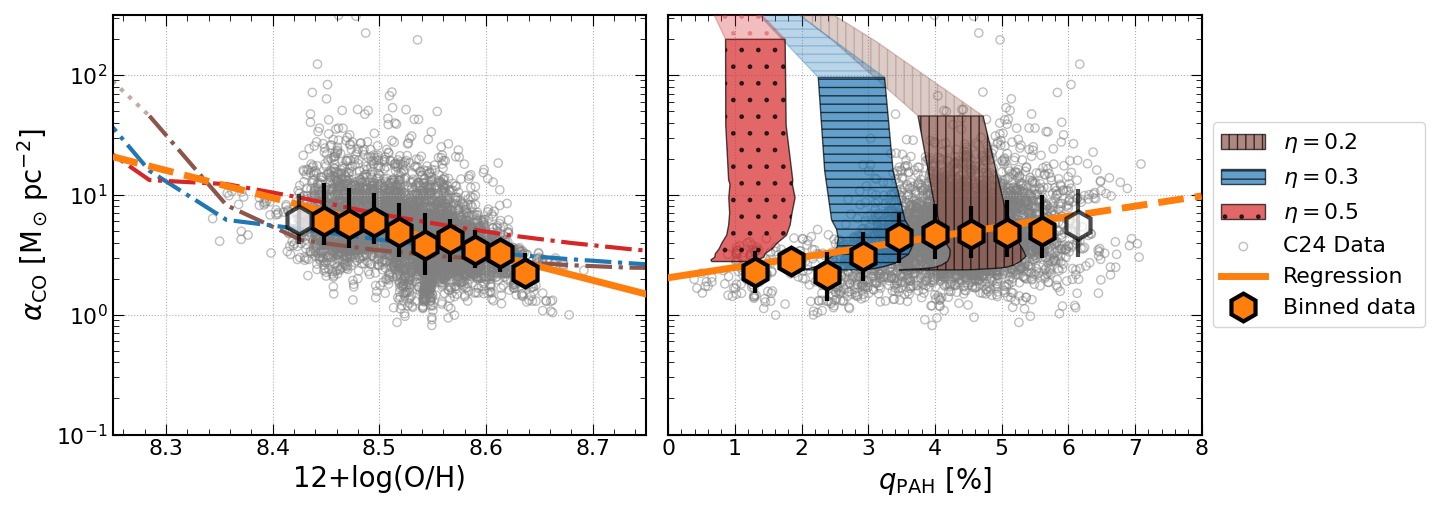}
    \caption{Relation between measured and modeled \aco and physical parameters, metallicity (left) and \qpah (right).
    The small empty circles show the distribution of pixel-by-pixel measurements.
    The orange hexagons show the binned data distribution, while the orange line shows the linear regression of pixel-by-pixel measurements.
    We show the model prediction from \citetalias{Hirashita23b_aCO} with different values of dense gas fraction in brown, blue and red lines (left panel) or shaded areas (right panel) as shown in the legend.
    }
    \label{fig:aCO-prop}
\end{figure*}

Here we discuss the dependence of \aco on grain size distribution. As mentioned at the beginning of this section, we use \qpah as the indicator for grain size distribution in this work. 
We show how \aco behaves with metallicity and \qpah in \autoref{fig:aCO-prop}. Both the observed and model-predicted \aco negatively correlate with metallicity, and the binned observed data falls in the model prediction range. The predicted \aco--metallicity relations at different $\eta$ do not differ significantly considering the scatter in observed data. Meanwhile, the large observed scatter of \aco at fixed metallicity is not simply explained by the variation in the grain size distribution, thus physical conditions other than dust evolution, such as gas temperature and velocity dispersion which affect the CO emissivity \citep[e.g.\ \citetalias{Hirashita23b_aCO} and ][]{Teng24_alphaCO_dv}, should play a role in producing the variation in \aco at a fixed metallicity. We refer readers to the reviews \citet{BOLATTO13} and \citet{SchinnererLeroy24_SFReview} for the impacts of non-dust evolution mechanisms on \aco.

For the dependence of \aco on \qpah, there is a considerable difference between observed and model-predicted data sets. The observed \aco positively correlates with \qpah. The Spearman's correlation coefficient ($\rho$) of \aco with \qpah is 0.21, which is weaker than the ones with metallicity ($\rho=-0.41$) and $r_\mathrm{g}/R_{25}$ ($\rho=0.28$), meaning that this correlation between \aco and \qpah is secondary compared to other correlations. 
Meanwhile, the model-predicted \qpah has no clear correlation with \aco. As shown in \autoref{sec:results:qpah}, \qpah is roughly constant once $\eta$ is set, while \aco still varies with metallicity.
Thus, although our model with $\eta =0.2$ broadly explains the \qpah--metallicity and \aco--metallicity relations, it does not seem to reproduce the observed \aco--\qpah relation.
In other words, the observed \aco--\qpah relation may be caused by mechanisms other than metal/dust enrichment; for example, by the local physical condition (radiation field, gas temperature, etc.).
Besides \qpah, \citetalias{Hirashita23b_aCO} predicted that \aco decreases with an increased presence of small dust grains, as stronger dust shielding occurs. However, we observe that \aco increases with \qpah. This could mean that \qpah does not work well as an indicator for size distribution, or that the observed \aco--\qpah relation is dominated by a mechanism irrelevant to dust evolution.

\begin{figure*}
    \centering
	\includegraphics[width=1.0\textwidth]{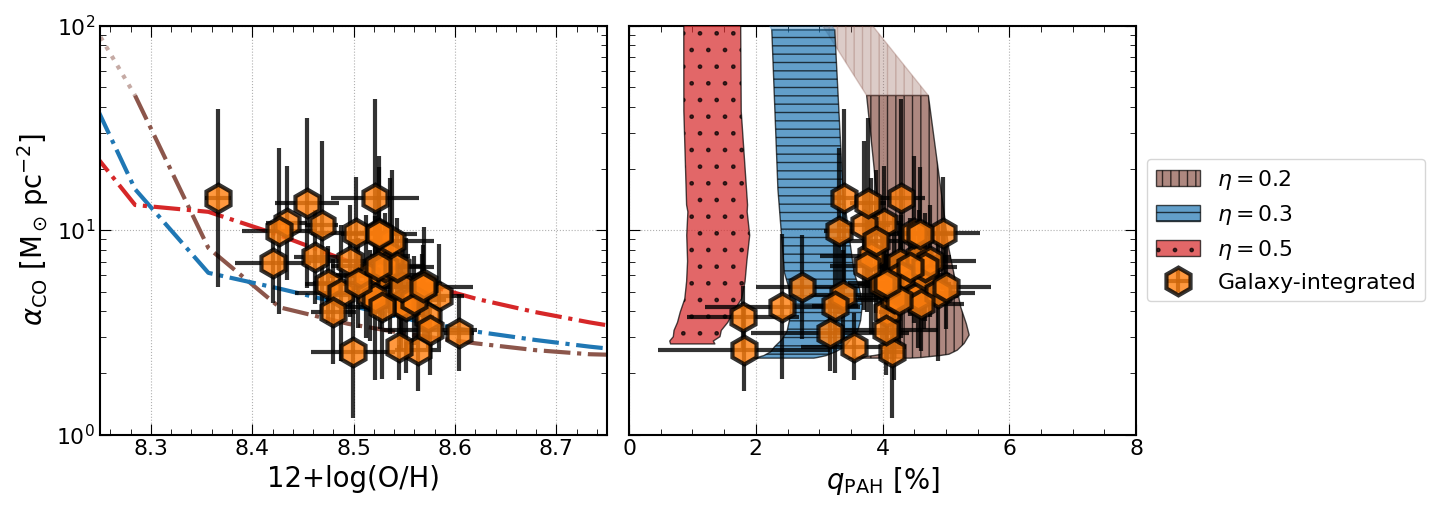}
    \caption{Measured and modelled \aco in terms of metallicity (left) and \qpah (right) for galaxy-integrated observations.
    Each orange hexagon represents one observed galaxy. The completeness threshold is not considered for the galaxy-integrated regression. The lines for the model predictions are the same as those shown in \autoref{fig:aCO-prop}.
    }\label{fig:aCO-prop_galaxy}
\end{figure*}

The above result implies the limitation in our model. Our evolution model of grain size distribution is based on a one-zone treatment where all physical conditions are set to be uniform; thus, it does not consider local physical conditions. 
Therefore, we also investigate the same relations as above for the galaxy-integrated properties, where local variations are averaged. This is shown in \autoref{fig:aCO-prop_galaxy}. For the observational data set, we take \Sigmad-weighted averaged \qpah, \Sigmagas-weighted \metal and $I_\mathrm{CO}$-weighted averaged \aco.

In the left panel of \autoref{fig:aCO-prop_galaxy}, we show that the galaxy-integrated \aco still negatively correlates with metallicity, which is consistent with \citetalias{Chiang24_alphaCO_main}. The models are also consistent with the observed \aco--metallicity relation.
In the right panel, we find that the observed \aco--\qpah pairs of most galaxies fall within the prediction of the $\eta=0.2$ model on the \aco--\qpah plane. A weak correlation between \aco and \qpah remains, which is not accounted for by our model. Nevertheless, our model successfully reproduces the overall galaxy-integrated \aco--\qpah--metallicity relations.

We do not obtain significant evidence that the grain size distribution affects \aco. The positive trend between \aco and \qpah is opposite to what is expected from the grain size distribution. Indeed, a negative correlation would be expected since large \qpah would mean more small grains which decrease \aco \citepalias{Hirashita23b_aCO}. The positive correlation likely comes from the negative correlation between \qpah and metallicity (or dust abundance); that is, if the metallicity (or dust abundance) is high, both \aco and \qpah are lowered (Sections \ref{sec:results:qpah} and \ref{sec:results:alphaCO}), producing a positive correlation between \aco and \qpah. Thus, the effect of dust enrichment (increase in dust abundance) is more prominent than the change in grain size distribution. However, this does not necessarily mean that the effect of grain size distribution is negligible. We further make an effort to extract the signature of grain size distribution in \autoref{sec:discussion:aco}.

\section {Discussion}\label{sec:discussion}

\subsection{Alternative indicator for grain size distribution}\label{sec:discussion:grains_size}

\begin{figure}
    \centering
	\includegraphics[width=\columnwidth]{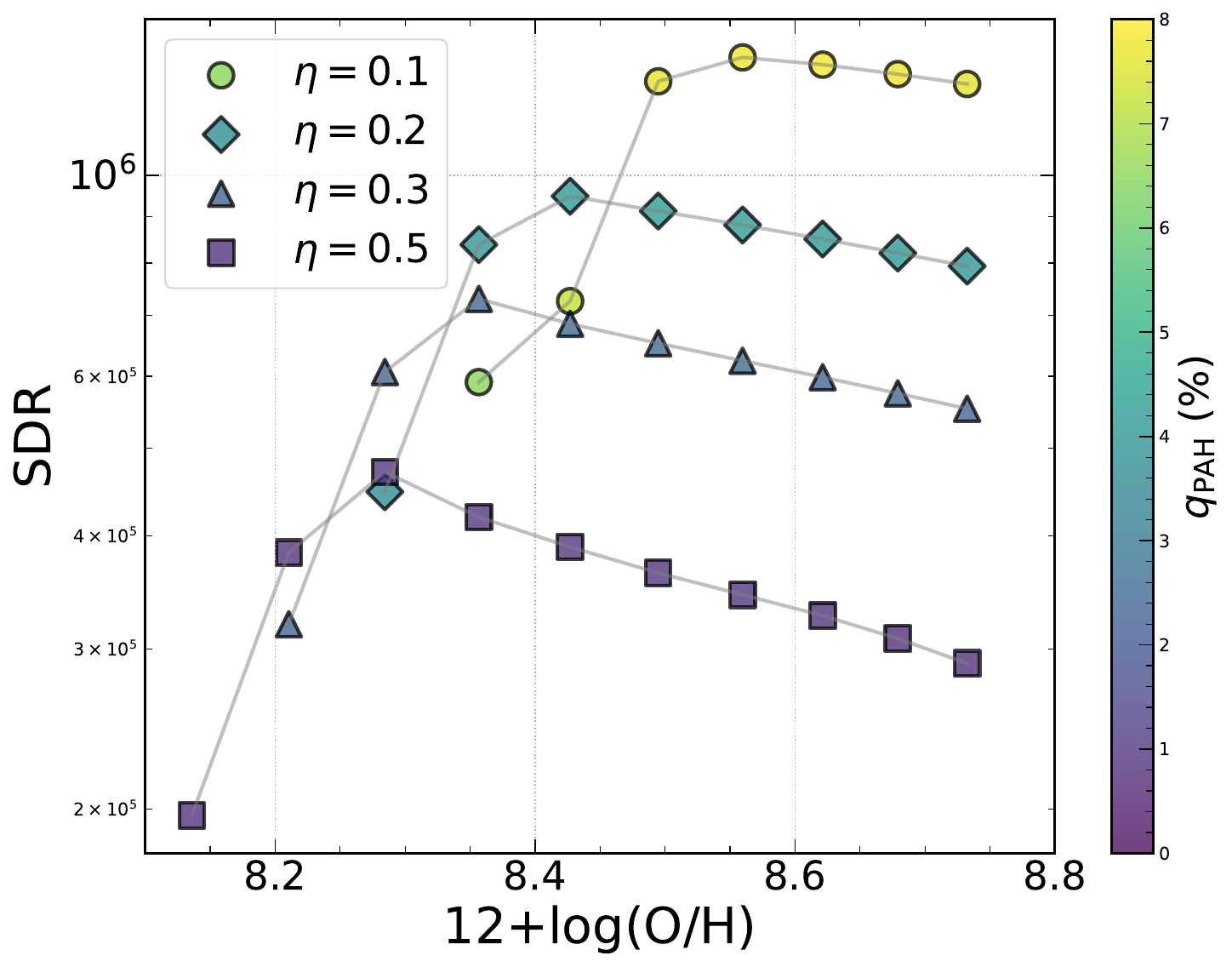}
    \caption{The model-predicted evolution of SDR with metallicity, colourized by \qpah. The symbols show tracks with different $\eta$ as indicated in the legend. As the colour scale indicates, \qpah traces the variation in SDR with $\eta$, but not the variation at fixed $\eta$.
    }
    \label{fig:SDR-qpah-metal_evolution}
\end{figure}

In \autoref{sec:results}, we used \qpah as the indicator of dust grain size distribution. This might not be the ultimate solution, as theoretically, \qpah only accounts for a portion of small carbonaceous grains.
Instead of \qpah, the theoretical work \citetalias{Hirashita23b_aCO} used the ratio between the total grain surface area and the total dust mass, SDR, as an indicator of grain size distribution, which better encompasses the overall size distribution. Higher SDR indicates that the grain size distribution is more biased towards smaller sizes.
\citetalias{Hirashita23b_aCO} suggested an empirical formula to correct \aco for the grain size distribution using SDR, which is expressed with the conversion from $X_\mathrm{CO}$ to \aco as:
\begin{align}
\aco=4.3\left(\frac{\mathcal{D}}{7\times 10^{-3}}\right)^{-2}\left(\frac{\mathrm{SDR}}{\mathrm{SDR}_0}\right)^{-0.5}\,\acoUnit,
\label{eq:alpha_CO_H23}
\end{align}
where $\mathcal{D}$ is the dust-to-gas ratio, and $\mathrm{SDR}_0=2.9\times 10^5~\mathrm{cm^2~g^{-1}}$ is a reference SDR.
We investigate whether we can find observational correspondence for SDR and then verify this prediction.

We first examine whether \qpah, a quantity shared by observation and model, can be used as an observational correspondence for SDR. In \autoref{fig:SDR-qpah-metal_evolution}, we show the relation between SDR, \qpah and metallicity.
To the first order, the values of both \qpah and SDR at high metallicity decrease with $\eta$, which results from the removal of small grains mainly due to coagulation. The variation in SDR is more significant.
At lower metallicity, \qpah is roughly constant at fixed $\eta$; meanwhile, SDR first increases with metallicity then decreases, spanning up to a factor of 2. In other words, we observe that, at fixed $\eta$, one \qpah value could map to multiple SDR values.
Moreover, the typical uncertainty for the observed \qpah in the \zzmgsherschel catalogue is $\sim 0.2\%$, which is similar to the predicted span of \qpah at fixed $\eta$. These facts make \qpah a suboptimal candidate for serving as the sole tracer for SDR.

We thus move on and examine whether the other quantities shared by the \citetalias{Hirashita23b_aCO} model and \citetalias{Chiang24_alphaCO_main} observation traces the change of SDR at fixed $\eta$. Since we assume fixed D/M in the observational data, there is practically only one shared observable left, which is metallicity.
As shown in \autoref{fig:SDR-qpah-metal_evolution}, at fixed $\eta$, each metallicity value only maps to one SDR value, which makes it a potential tracer for SDR.
On the other hand, metallicity does not catch how SDR varies with $\eta$, which means that we need both \qpah and metallicity to infer SDR.

\begin{figure}
    \centering
	\includegraphics[width=\columnwidth]{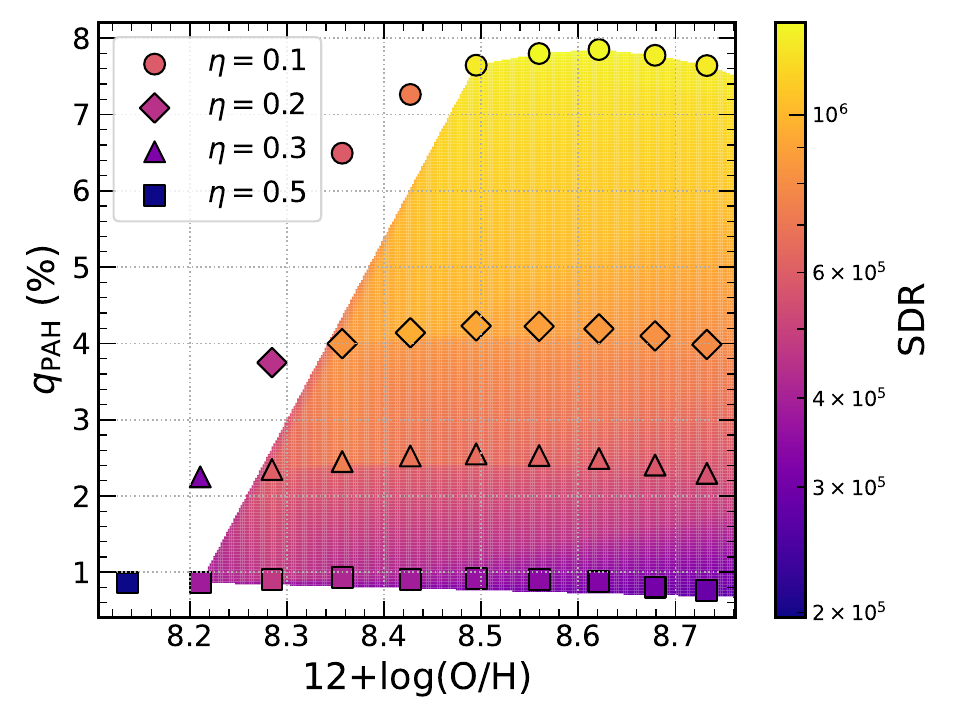}
    \caption{Relation among SDR, \qpah and metallicity. The symbols show the model-predicted dataset, and the shaded region presents the 2-dimensional interpolation of SDR. We drop the low-metallicity points because we cannot reach a smooth interpolation with them. 
    }
    \label{fig:qpah-metal-SDR}
\end{figure}

Based on the above arguments, we expect that SDR could be expressed by a combination of \qpah and metallicity. Thus, we interpolate SDR on the \qpah--metallicity 2-dimensional plane.
In the interpolation, we used model-predicted values at various $\eta$, omitting the data points at early evolution stages (low metallicity) for smoothness.
The resulting interpolation is shown in \autoref{fig:qpah-metal-SDR}, and is used to infer SDR values for each observed pixel from their \qpah and metallicity measurements.

\subsection{Variation in \texorpdfstring{\aco}{the conversion factor} with SDR}\label{sec:discussion:aco}

\begin{figure*}
    \centering
	\includegraphics[width=1.0\textwidth]{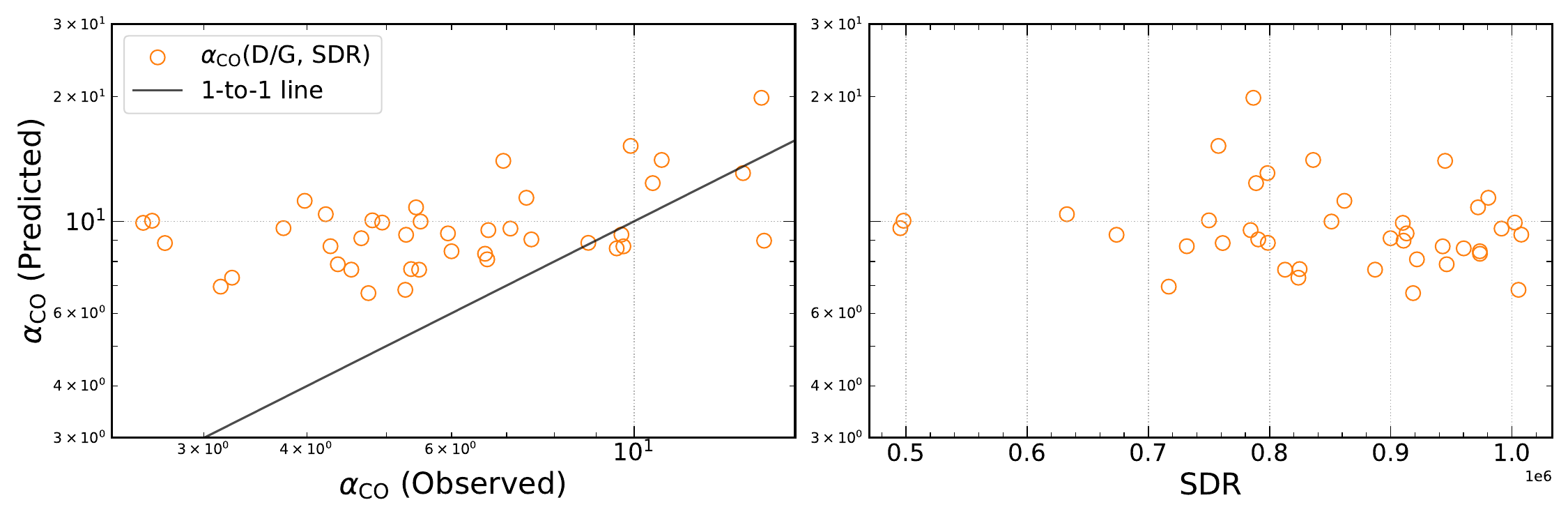}
    \caption{\aco and grain size distribution. Left: \aco predicted as a function of D/G and SDR (equation\,18 in \citetalias{Hirashita23b_aCO}) versus the galaxy-integrated observations. Right: The predicted \aco versus SDR. There is no obvious trend between \aco and SDR, suggesting that the dependence of \aco on SDR is only secondary.}\label{fig:aCO-pred-obs_galaxy}
\end{figure*}

In \autoref{fig:aCO-pred-obs_galaxy} (left panel), we show the \aco values predicted by equation (\ref{eq:alpha_CO_H23}), which are compared with the observed \aco in each galaxy. The SDR value is from the interpolation described in \autoref{sec:discussion:grains_size}.
We find a systematic trend of overpredicting \aco at low observed \aco. In other words, the \aco values from \citetalias{Hirashita23b_aCO} models have a smaller dynamic range than those from the observations, implying that physical factors other than dust evolution also influence \aco strongly. The overestimating trend at low \aco could be explained by the \aco decline due to increased CO emissivity \citep[e.g.][]{Hayashi19,Teng24_alphaCO_dv,Chiang24_alphaCO_main,SchinnererLeroy24_SFReview}, which is not taken into account in the \citetalias{Hirashita23b_aCO} model.
Recent studies on the molecular clouds in the MW and Large Magellanic Cloud also showed that \aco could vary by almost an order of magnitude at fixed metallicity \citep{Kohno24,Kohno24b,Li24_XCO}, 
This is consistent with the arguments in \autoref{sec:results:alphaCO} based on the \aco--\qpah relation.

To further understand the prediction with the dust evolution mechanisms only, we examine the components in equation (\ref{eq:alpha_CO_H23}): the total dust abundance (D/G) term and the grain size distribution (SDR) term. The D/G term is broadly consistent with literature values \citep[e.g.][]{SCHRUBA12,BOLATTO13,HUNT15,ACCURSO17,SchinnererLeroy24_SFReview}; thus, we focus on the SDR term.
In order to examine the effect of SDR, we first show in \autoref{fig:aCO-pred-obs_galaxy} (right panel) the \aco values predicted from equation (\ref{eq:alpha_CO_H23}) in terms of SDR. We do not find any significant trend. This indicates that the impact of grain size distribution on \aco is secondary compared to other mechanisms among the sample galaxies. However, the above result does not indicate that the grain size distribution is unimportant.
It is fair to say that the dynamic range of SDR is small compared with other quantities affecting \aco at galaxy-integrated scales because the grain size distribution is converged to a functional shape determined by the balance between coagulation and shattering as we argued in \autoref{sec:results:qpah}.

\subsection{Alternative explanations of high-metallicity \texorpdfstring{\qpah}{PAH faction} decline}\label{sec:discussion:qpah_decline}

In \autoref{sec:results}, we focused on coagulation in the discussion on PAH destruction mechanisms, especially at high metallicity, and neglected other possibilities, e.g. photodestruction of PAHs in hard radiation fields or weaker PAH emission with softer radiation in the bulge. We will discuss these two cases here.

\subsubsection{Photodestruction of PAHs}\label{sec:discussion:qpah_decline:photodestruction}
Since the chemical structure of astronomical PAHs is not unique, we make a rough calculation of the energy required to dissociate a single bond in a benzene ring, the building block of PAHs.
A rough calculation of the dissociation energy includes the C--C bond, the C=C bond and the resonance energy. The average bond energy of C--C and C=C bonds are 347 and 614~kJ~mol$^{-1}$, respectively. The resonance energy of a benzene ring is 150~kJ~mol$^{-1}$. Thus, to photodissociate a single bond in a benzene ring, a photon with an energy of at least 5.27~eV ($\lambda \sim 235$~nm) is required. \citet{Kislov04_Benzene} calculated the probability of various pathways of photodissociation of benzene at different wavelengths. They found that the dissociation of H atoms starts to occur at $\lambda \lesssim 193$~nm. They also predicted that the dissociation of C atoms might be observable at $\lambda \leq 157$~nm. Thus, it seems possible to observe the photodestruction of PAHs under a hard radiation field. \cite{Murga19} calculated the PAH photodestruction time-scale as a function of ISRF strength. According to their calculation, the ISRFs in our dataset ($\log\Ubar \lesssim 1.5$) would predict PAH destruction of $\gtrsim 10^9~\mathrm{yr}$, which is less efficient than the reference coagulation time-scale of $\gtrsim 10^7~\mathrm{yr}$ \citep{Hirashita_Yan09_coagulation}. Thus although photodestruction of PAHs is possible in our sample, it is expected to be less efficient than the processes included in our models.

Besides theoretical works, previous observations have found evidence of PAHs being destroyed by strong radiation fields. For example, \citet{Chastenet19_MCs} showed that the destruction of PAHs, indicated by local \qpah relative to the average \qpah in each of their sample galaxies (the Magellanic Clouds), strongly correlates with the surface brightness of H$\alpha$ at 10~pc scale. They also showed that in the diffuse neutral medium, it is not clear whether the ISRF strength impacts the destruction of PAHs. Studies with \textit{Spitzer} IRAC photometry \citep{Khramtsova13_PAH} and recent (sub-)cloud scale studies with JWST also showed lowered \qpah in \HII regions \citep{Chastenet23_PAH_Variation,Egorov23}. 

Meanwhile, this work examines $\sim$kpc scales, focusing on neutral gas-dominated environments. Therefore, we need to carefully interpret sub-cloud scale findings in \HII regions for our analysis. \citet{Sutter24_PAH} used PHANGS-JWST observations and conducted several tests about how PAH band ratio ($R_\mathrm{PAH}\equiv\mathrm{[F770W+F1130W]/F2100W}$, which traces \qpah) varies depending on averaging methods.
First, they confirmed that $R_\mathrm{PAH}$ is systematically lower in \HII regions compared to diffuse regions at 10--50~pc scales. However, this effect is only significant when $R_\mathrm{PAH}$ is calculated as H$\alpha$-weighted average. When they took simple averages over larger regions, there is no significant difference in $R_\mathrm{PAH}$ between diffuse and \HII regions, indicating that the photodestruction effect is easily diluted when working at coarser resolution.
They then investigated whether $R_\mathrm{PAH}$ is affected by the presence of \HII regions in kpc-scale cells, and they did not find a significant trend of $R_\mathrm{PAH}$ with the percentage of pixels identified as \HII regions.
In summary, photodestruction of PAHs is theoretically possible and effective on much smaller scales than our adopted resolution ($\sim$kpc). Thus, it is unlikely that the decreasing trend of \qpah with metallicity presented in \autoref{sec:results} is due to photodestruction.


\subsubsection{Bulge correction for \texorpdfstring{\qpah}{PAH fraction}}\label{sec:discussion:qpah_decline:bulge}

The rate at which starlight heats dust differs with the starlight spectrum at a fixed mean ISRF strength (\Ubar). This effect impacts small grains more than large grains, causing possible misestimate of \qpah from the SED fitting when the starlight spectrum is different from the model assumption, e.g.\ when \Ubar is dominated by the older stellar population in the bulge \citep{DRAINE14,Whitcomb24PAH}. \citet{DRAINE14} provided a detailed strategy for correcting the underestimated \qpah when \Ubar consists of the bulge ISRF ($U_\mathrm{bulge}$) and the disc ISRF ($U_\mathrm{disc}$). However, the derivation of $U_\mathrm{bulge}$ requires robust measurements of bulge core radius for all our sample galaxies, which is beyond the scope of this work. Thus, we do not include this correction in our fiducial analysis.

\begin{figure*}
    \centering
	\includegraphics[width=\textwidth]{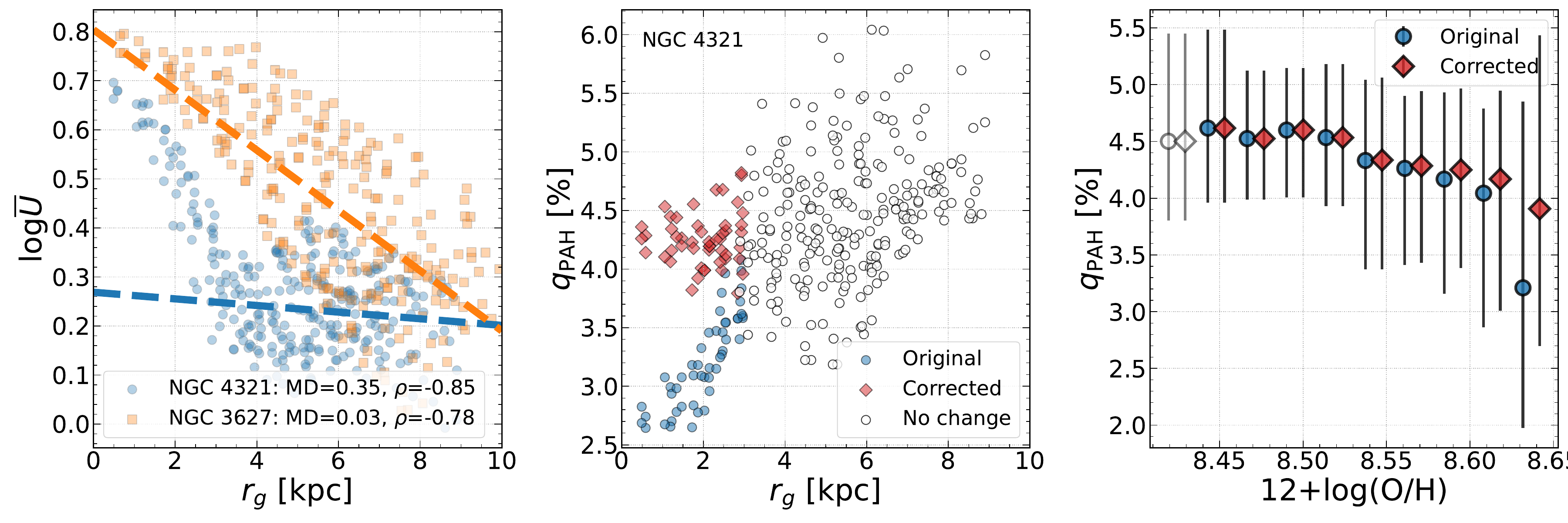}
    \caption{
    Left: Pixel-by-pixel \Ubar as a function of $r_\mathrm{g}$ in two example galaxies (orange circles and blue squares for NGC 4321 and NGC 3627, respectively). The dashed lines show the exponential disc fits ($U_\mathrm{disc}$) for each galaxy with corresponding colours. In NGC 4321, we could extract $U_\mathrm{bulge}$, while in NGC 3627, we could not extract $U_\mathrm{bulge}$ because of the small deviation between \Ubar and the extrapolated disc component in the inner galaxy.
    Middle: Pixel-by-pixel corrected \qpah compared to original \qpah as a function of $r_g$ in NGC 4321. The empty circles show where \qpah remains the same.
    Right: Binned corrected \qpah compared to original \qpah as a function of metallicity. The errorbar shows the 16th and 84th percentiles in each bin. The empty symbols present data outside the completeness cut.
    }
    \label{fig:bulge_correction}
\end{figure*}

Here, we use a simplified strategy to estimate the impact from $U_\mathrm{bulge}$, inspired by the ``scale-down'' strategy in \citet{DRAINE14} and the ``flat $U_\mathrm{disc}$'' strategy in \citet{Whitcomb24PAH}. The key assumptions in our strategy are:
(a) $\Ubar = U_\mathrm{bulge} + U_\mathrm{disc}$, where $U_\mathrm{disc}$ dominates beyond a certain galactocentric distance ($r_\mathrm{g}$); (b) in each galaxy, $U_\mathrm{disc}(r_\mathrm{g})\propto \mathrm{exp}(-r_\mathrm{g}/r_\mathrm{d})$, where $r_\mathrm{d}$ is a galaxy-dependent scale length of the disc.

In each galaxy, we first fit \Ubar to $U_\mathrm{disc}(r_g)$ at $r_g > 3~\mathrm{kpc}$. We then extrapolate the fitted $U_\mathrm{disc}(r_g)$ to the inner galaxy, and calculate $U_\mathrm{bulge} = \Ubar - U_\mathrm{disc}$. We take galaxies that meet the following criteria to have a significant bulge component for our analysis: (a) the median deviation (MD) between log\Ubar and $\log U_\mathrm{disc}$ in the inner 2~kpc exceeds 0.1~dex, and (b) the Spearman's correlation ($\rho$) between $U_\mathrm{bulge}$ and $r_g$ in the inner 2~kpc is stronger than $-0.5$, with a $p$-value below 0.05. We end up with 6 galaxies satisfying the above criteria: IC 342, NGC 3184, NGC 4051, NGC 4321, NGC 5457 and NGC 6946. Lastly, we adopt equation~(21) in \citet{DRAINE14} to calculate \textit{corrected} \qpah. Overall, the median of increase in \qpah after the correction is 1.12 per cent.\footnote{This is calculated only for the pixels affected by the correction.}

We show some example galaxies and the resulting \textit{corrected} \qpah in \autoref{fig:bulge_correction}. In the left panel, we show the \Ubar (symbols) and fitted $U_\mathrm{disc}$ (dashed lines) in two galaxies. NGC 4321 satisfies both criteria, while NGC 3627 does not show large enough MD to extract $U_\mathrm{bulge}$. In the middle panel of \autoref{fig:bulge_correction}, we show the \textit{corrected} and original \qpah in NGC 4321 as an example. This correction impacts \qpah at $r_\mathrm{g} \lesssim 3~\mathrm{kpc}$ depending on how \Ubar behaves with $r_g$ and where we fit $U_\mathrm{disc}(r_g)$. The corrected \qpah significantly removes the decline of \qpah in the inner galaxy shown with the original measurements. The median increase in \qpah is 0.64 per cent for NGC 4321. In the right panel of \autoref{fig:bulge_correction}, we show how \textit{corrected} and original \qpah vary with metallicity for all galaxies, in a manner similar to \autoref{fig:qpah-metal} (fiducial sample). In the highest metallicity bin, \qpah rises by almost 1 per cent after the correction, while the other bins are barely affected.
In summary, the correction for the bulge radiation field increases the \qpah values only in the innermost (highest-metallicity) part of galaxies. It weakens the negative trend of \qpah with metallicity but does not remove it. Thus, we conclude that the different stellar populations in high-metallicity regions do not significantly affect the observationally derived \qpah--metallicity relation.

\section{Summary}\label{sec:summary}
We investigate how grain size distribution affects the CO-to-H$_2$ conversion factor (\aco) and how the PAH fraction (\qpah), which is used as an indicator of grain size distribution, evolves with local environments with both observations and models. We adopt the \citetalias{Chiang24_alphaCO_main} measurements of \aco, metallicity, and \qpah at 2~kpc resolution in 42 nearby galaxies. The \aco is derived from measured $W_\mathrm{CO}$, \Sigmad, \Sigmaatom and metallicity assuming $\mathrm{D/M}=0.48$. The dust properties are derived from the IR SED fitting with the \citet{DRAINE07} model as part of the \zzmgsherschel work \citep{Chastenet24_Herschel}. We utilize the \citetalias{Hirashita23b_aCO} analytical model that calculates \aco in a manner consistent with the evolution of dust abundance and grain size distribution. It is capable of predicting \aco, D/G, metallicity, \qpah and SDR at each evolutionary stage.

We find a weak anti-correlation between the observed \qpah and metallicity, especially at $\metal > 8.4$. This anti-correlation is stronger in CO-bright environments. Meanwhile, the \citetalias{Hirashita23b_aCO} models predict a roughly constant \qpah at mid- to high-metallicity at fixed $\eta$ (dense gas fraction), which is more consistent with the \textit{broad} sample (omitting S/N cuts).
On the other hand, the equilibrium \qpah is set by the balance between shattering and coagulation. Lower $\eta$ values yield higher equilibrium \qpah due to weaker coagulation (or stronger shattering), and the $\eta = 0.2$ prediction best matches our observations.
The enPAH model, which assumes no coagulation for PAHs, does not align with observed trends. This suggests an important role of coagulation in reproducing the observed \qpah--metallicity relation.

We discuss how dust properties could impact \aco. We first compare the observed and modelled \aco and examine the dependence on metallicity and \qpah. The observations and the model predictions show similar anti-correlation between \aco and metallicity, while the observations have a larger span of \aco at fixed metallicity than explained by the models. Meanwhile, the observations and the models show a considerable difference in the \aco--\qpah relation. The observed \aco shows a positive correlation with \qpah, whereas the model-predicted \aco lacks a clear correlation with \qpah. On the other hand, galaxy-integrated observations show consistent results with the predictions, indicating that the discrepancy we show in the pixel-by-pixel analysis is likely due to the limitation arising from the one-zone treatment in the model.

We also investigate how \aco depends on SDR, the ratio between the total grain surface area and the total dust mass, which is an alternative tracer of grain size distribution adopted by \citetalias{Hirashita23b_aCO}. 
We first examine the relation between SDR and observed quantities and find that \qpah traces the variation of SDR with $\eta$. However, because of the uncertainty level and the lack of a one-to-one mapping, \qpah does not trace the evolution of SDR at fixed $\eta$. Meanwhile, we find a one-to-one mapping from metallicity to SDR at fixed $\eta$. With the combination of \qpah and metallicity, we build a 2-dimensional interpolation map to assign SDR values to observational data.
We find that the \aco predicted from SDR and metallicity with the formula derived in \citetalias{Hirashita23b_aCO} is more consistent with the observations at larger \aco values.
While SDR affects \aco, the impact of the SDR term to \aco value is secondary compared to other physical conditions in the ISM, e.g.\ the \aco decline due to increased CO emissivity.

We discuss the possible reasons for the \qpah decline at high metallicity besides coagulation. PAHs could be destroyed by hard radiation fields in \HII regions. However, this effect is likely unobservable at our 2~kpc resolution. The bulge ISRF correction could raise \qpah at near solar metallicity by almost 1 per cent from our estimation. However, this effect alone does not explain the negative \qpah--metallicity correlation. Given that the above two mechanisms do not completely explain this negative correlation, coagulation remains a viable process that naturally explains the decrease of \qpah with metallicity.

\section*{Acknowledgements}
We thank the anonymous referee for their feedback, which has helped to improve this work.
We thank Cory Whitcomb for the useful discussion on the bulge ISRF correction.
IC and HH thank the National Science and Technology Council for support through grant 111-2112-M-001-038-MY3, and the Academia Sinica for Investigator Award AS-IA-109-M02 (PI: Hiroyuki
Hirashita).
JC acknowledges funding from the Belgian Science Policy Office (BELSPO) through the PRODEX project ``JWST/MIRI Science exploitation'' (C4000142239).
EWK acknowledges support from the Smithsonian Institution as a Submillimeter Array (SMA) Fellow.

This work uses observations made with ESA \textit{Herschel} Space Observatory. \textit{Herschel} is an ESA space observatory with science instruments provided by European-led Principal Investigator consortia and with important participation from NASA. The \textit{Herschel} spacecraft was designed, built, tested, and launched under a contract to ESA managed by the \textit{Herschel}/Planck Project team by an industrial consortium under the overall responsibility of the prime contractor Thales Alenia Space (Cannes), and including Astrium (Friedrichshafen) responsible for the payload module and for system testing at spacecraft level, Thales Alenia Space (Turin) responsible for the service module, and Astrium (Toulouse) responsible for the telescope, with in excess of a hundred subcontractors.

This paper makes use of the VLA data with project codes 14A-468, 14B-396, 16A-275 and 17A-073, which has been processed as part of the EveryTHINGS survey.
This paper makes use of the VLA data with legacy ID AU157, which has been processed in the PHANGS--VLA survey.
The National Radio Astronomy Observatory is a facility of the National Science Foundation operated under cooperative agreement by Associated Universities, Inc. 
This publication makes use of data products from the Wide-field Infrared Survey Explorer, which is a joint project of the University of California, Los Angeles, and the Jet Propulsion Laboratory/California Institute of Technology, funded by the National Aeronautics and Space Administration.

This paper makes use of the following ALMA data, which have been processed as part of the PHANGS--ALMA CO(2--1) survey: \linebreak
ADS/JAO.ALMA\#2012.1.00650.S, \linebreak 
ADS/JAO.ALMA\#2015.1.00782.S, \linebreak 
ADS/JAO.ALMA\#2018.1.01321.S, \linebreak 
ADS/JAO.ALMA\#2018.1.01651.S. \linebreak 
ALMA is a partnership of ESO (representing its member states), NSF (USA) and NINS (Japan), together with NRC (Canada), MOST and ASIAA (Taiwan), and KASI (Republic of Korea), in cooperation with the Republic of Chile. The Joint ALMA Observatory is operated by ESO, AUI/NRAO and NAOJ.

This research made use of Astropy,\footnote{http://www.astropy.org} a community-developed core Python package for Astronomy \citep{ASTROPY13,Astropy18,Astropy22}. 
This research has made use of NASA's Astrophysics Data System Bibliographic Services. 
We acknowledge the usage of the HyperLeda database (http://leda.univ-lyon1.fr). 
This research has made use of the NASA/IPAC Extragalactic Database (NED), which is funded by the National Aeronautics and Space Administration and operated by the California Institute of Technology.

\section*{Data Availability}

Data related to this publication and its figures are available on reasonable request from the corresponding author. 



\bibliographystyle{mnras}
\bibliography{references} 




\appendix

\section{Galaxy samples}\label{app:samples}
We present the properties of galaxies studies in this work and the references for adopted data in \autoref{tab:samples}.

\begin{table*}
\caption{Galaxy Sample.}\label{tab:samples}
\begin{tabular}{cccccccccccc}
    \hline
    \hline
    Galaxy & Dist. & $i$ & P.A. & $R_{25}$\hspace{0em} & $R_e$\hspace{0em} & $\log(M_\star)$ & Type & \coone & \cotwo & 	\HI 21~cm Ref & 12+log(O/H) Ref \\
    & [Mpc] & [\arcdeg] & [\arcdeg] & [kpc] & [kpc] & [M$_\odot$] &  &  &  &  &  \\
    (1) & (2) & (3) & (4) & (5) & (6) & (7) & (8) & (9) & (10) & (11) & (12) \\
    \hline
    IC~342 & 3.5 & 31.0 & 42.0 & 10.1 & 4.4 & 10.2 & 5 & CO Atlas & \nodata & EveryTHINGS & $f.$ \\
    NGC~253 & 3.7 & 75.0 & 52.5 & 14.4 & 4.7 & 10.5 & 5 & CO Atlas & PHANGS-ALMA & $c.$ & $g.$ \\
    NGC~300 & 2.1 & 39.8 & 114.3 & 5.9 & 2.0 & 9.3 & 6 & \nodata & PHANGS-ALMA & $d.$ & $h.$ \\
    NGC~598 & 0.9 & 55.0 & 201.0 & 8.1 & 2.4 & 9.4 & 5 & \nodata & $a.$ & $e.$ & $h.$ \\
    NGC~628 & 9.8 & 8.9 & 20.7 & 14.1 & 3.9 & 10.2 & 5 & COMING & PHANGS-ALMA & THINGS & PHANGS-MUSE \\
    NGC~925* & 9.2 & 66.0 & 287.0 & 14.3 & 4.5 & 9.8 & 6 & \nodata & HERACLES & THINGS & $h.$ \\
    NGC~2403* & 3.2 & 63.0 & 124.0 & 9.3 & 2.4 & 9.6 & 5 & \nodata & HERACLES & THINGS & $h.$ \\
    NGC~2841 & 14.1 & 74.0 & 153.0 & 14.2 & 5.4 & 10.9 & 3 & COMING & \nodata & THINGS & $g.$ \\
    NGC~2976* & 3.6 & 65.0 & 335.0 & 3.0 & 1.3 & 9.1 & 5 & COMING & HERACLES & THINGS & $g.$ \\
    NGC~3184 & 12.6 & 16.0 & 179.0 & 13.6 & 5.3 & 10.3 & 5 & CO Atlas & HERACLES & THINGS & $h.$ \\
    NGC~3198 & 13.8 & 72.0 & 215.0 & 13.0 & 5.0 & 10.0 & 5 & COMING & HERACLES & THINGS & $g.$ \\
    NGC~3351 & 10.0 & 45.1 & 193.2 & 10.5 & 3.1 & 10.3 & 3 & CO Atlas & PHANGS-ALMA & THINGS & PHANGS-MUSE \\
    NGC~3521 & 13.2 & 68.8 & 343.0 & 16.0 & 3.9 & 11.0 & 3 & CO Atlas & PHANGS-ALMA & THINGS & $g.$ \\
    NGC~3596 & 11.3 & 25.1 & 78.4 & 6.0 & 1.6 & 9.5 & 5 & \nodata & PHANGS-ALMA & EveryTHINGS & $g.$ \\
    NGC~3621 & 7.1 & 65.8 & 343.8 & 9.9 & 2.7 & 10.0 & 6 & \nodata & PHANGS-ALMA & THINGS & $h.$ \\
    NGC~3627 & 11.3 & 57.3 & 173.1 & 16.9 & 3.6 & 10.7 & 3 & CO Atlas & PHANGS-ALMA & THINGS & PHANGS-MUSE \\
    NGC~3631 & 18.0 & 32.4 & -65.6 & 9.7 & 2.9 & 10.2 & 5 & CO Atlas & $b.$ & EveryTHINGS & $g.$ \\
    NGC~3938 & 17.1 & 14.0 & 195.0 & 13.4 & 3.7 & 10.3 & 5 & COMING & HERACLES & HERACLES-VLA & $g.$ \\
    NGC~3953 & 17.1 & 61.5 & 12.5 & 15.2 & 5.3 & 10.6 & 4 & \nodata & $b.$ & EveryTHINGS & $g.$ \\
    NGC~4030 & 19.0 & 27.4 & 28.7 & 10.5 & 2.1 & 10.6 & 4 & COMING & \nodata & EveryTHINGS & $g.$ \\
    NGC~4051 & 17.1 & 43.4 & -54.8 & 14.7 & 3.7 & 10.3 & 3 & CO Atlas & $b.$ & EveryTHINGS & $g.$ \\
    NGC~4207 & 15.8 & 64.5 & 121.9 & 3.5 & 1.4 & 9.6 & 7 & \nodata & PHANGS-ALMA & PHANGS-VLA & $g.$ \\
    NGC~4254 & 13.1 & 34.4 & 68.1 & 9.6 & 2.4 & 10.3 & 5 & CO Atlas & PHANGS-ALMA & HERACLES-VLA & PHANGS-MUSE \\
    NGC~4258 & 7.6 & 68.3 & 150.0 & 18.8 & 5.9 & 10.7 & 4 & COMING & \nodata & HALOGAS & $h.$ \\
    NGC~4321 & 15.2 & 38.5 & 156.2 & 13.5 & 5.5 & 10.7 & 3 & CO Atlas & PHANGS-ALMA & HERACLES-VLA & PHANGS-MUSE \\
    NGC~4450 & 16.8 & 48.5 & -6.3 & 13.3 & 4.3 & 10.7 & 2 & \nodata & $b.$ & EveryTHINGS & $g.$ \\
    NGC~4496A* & 14.9 & 53.8 & 51.1 & 7.3 & 3.0 & 9.6 & 6 & \nodata & PHANGS-ALMA & EveryTHINGS & $g.$ \\
    NGC~4501 & 16.8 & 60.1 & -37.8 & 21.1 & 5.2 & 11.0 & 3 & CO Atlas & \nodata & EveryTHINGS & $g.$ \\
    NGC~4536 & 16.2 & 66.0 & 305.6 & 16.7 & 4.4 & 10.2 & 3 & CO Atlas & PHANGS-ALMA & HERACLES-VLA & $g.$ \\
    NGC~4569 & 15.8 & 70.0 & 18.0 & 21.0 & 5.9 & 10.8 & 2 & CO Atlas & PHANGS-ALMA & HERACLES-VLA & $g.$ \\
    NGC~4625 & 11.8 & 47.0 & 330.0 & 2.4 & 1.2 & 9.1 & 9 & \nodata & HERACLES & HERACLES-VLA & $h.$ \\
    NGC~4651 & 16.8 & 50.1 & 73.8 & 9.5 & 2.4 & 10.3 & 5 & \nodata & $b.$ & EveryTHINGS & $h.$ \\
    NGC~4689 & 15.0 & 38.7 & 164.1 & 8.3 & 4.7 & 10.1 & 5 & CO Atlas & PHANGS-ALMA & EveryTHINGS & $g.$ \\
    NGC~4725 & 12.4 & 54.0 & 36.0 & 17.5 & 6.0 & 10.8 & 1 & \nodata & HERACLES & HERACLES-VLA & $g.$ \\
    NGC~4736 & 4.4 & 41.0 & 296.0 & 5.0 & 0.8 & 10.3 & 1 & CO Atlas & HERACLES & THINGS & $g.$ \\
    NGC~4941 & 15.0 & 53.4 & 202.2 & 7.3 & 3.4 & 10.1 & 1 & \nodata & PHANGS-ALMA & EveryTHINGS & $g.$ \\
    NGC~5055 & 9.0 & 59.0 & 102.0 & 15.5 & 4.2 & 10.7 & 4 & CO Atlas & HERACLES & THINGS & $g.$ \\
    NGC~5248 & 14.9 & 47.4 & 109.2 & 8.8 & 3.2 & 10.3 & 3 & CO Atlas & PHANGS-ALMA & PHANGS-VLA & $g.$ \\
    NGC~5457 & 6.7 & 18.0 & 39.0 & 23.4 & 13.5 & 10.3 & 5 & CO Atlas & HERACLES & THINGS & $h.$ \\
    NGC~6946 & 7.3 & 33.0 & 243.0 & 12.1 & 4.4 & 10.5 & 5 & CO Atlas & HERACLES & THINGS & $h.$ \\
    NGC~7331 & 14.7 & 76.0 & 168.0 & 19.8 & 3.7 & 11.0 & 4 & COMING & HERACLES & THINGS & $g.$ \\
    NGC~7793* & 3.6 & 50.0 & 290.0 & 5.4 & 1.9 & 9.3 & 6 & \nodata & PHANGS-ALMA & THINGS & $h.$ \\
    \hline
\end{tabular}
\begin{tablenotes}
    \item \small {\bf Notes:} (1) Name of galaxies. We mark the galaxies not presented in \citetalias{Chiang24_alphaCO_main} with ``*''; (2) Distance \citep[from EDD][]{dist_galbase_EDD_TULLY09}; (3-4) inclination angle and position angle \citep{1999ApJ...523..136S,DEBLOK08,LEROY09,MUNOZMATEOS09,2009ApJ...702..277M,2013ApJ...774..126M,MAKAROV14,LANGMEIDT_2020ApJ...897..122L}; (5) isophotal radius \citep{MAKAROV14}; (6) effective radius \citep{Leroy21_PHANGS-ALMA_CO}; (7) logarithmic global stellar mass \citep{LEROY19}; (8) numerical Hubble stage T; 
    (9) References of CO~$J=1\to0$ observations (``\nodata'' means no CO~$J=1\to0$ data adopted in this work):
    CO Atlas \citet{Kuno07_NRO_CO}; COMING \citep{Sorai19_COMING};
    (10) References of CO~$J=2\to1$ observations (``\nodata'' means no CO~$J=2\to1$ data adopted in this work):
    HERACLES \citet{LEROY09}; PHANGS-ALMA \citep{Leroy21_PHANGS-ALMA_CO}; 
    $a.$ M33 data from \citet{GRATIER10,DRUARD14};
    $b.$ New HERA data \citep[P.I.: A. Schruba; presented in][]{Leroy21_CO_Line_Ratios};
    (11) References of \textsc{Hi} observations:
    THINGS \citep{WALTER08}; HALOGAS \citep{Heald11_HALOGAS}; HERACLES-VLA \citep{SCHRUBA11}; PHANGS-VLA (P.I. D. Utomo; I. Chiang et al. in preparation); EveryTHINGS (P.I. K. M. Sandstrom; presented in \citetalias{Chiang24_alphaCO_main});
    $c.$ \citet{Puche91};
    $d.$ \citet{Puche90};
    $e.$ \citet{KOCH18};
    (12) References of \metal measurement:
    PHANGS-MUSE \citep{Emsellem22_PHANGS-MUSE,Santoro22}; 
    $f.$ private communication with K. Kreckel \citep[see][]{Chiang21};
    $g.$ using the empirical formula described in \citetalias{Chiang24_alphaCO_main};
    $h.$ data from \citet{Zurita21} compilation.
\end{tablenotes}
\end{table*}


\bsp	
\label{lastpage}
\end{document}